\documentclass[twocolumn, 11pt]{article}
\usepackage{amsmath}
\usepackage{amsfonts}
\usepackage{graphicx}
\usepackage{hyperref}
\usepackage{url}
\usepackage{xcolor}
\usepackage{microtype}
\usepackage{booktabs}
\usepackage{fancyhdr}
\usepackage{lipsum}
\usepackage{siunitx}
\usepackage[T1]{fontenc}
\usepackage[a4paper, left=2.1cm, right=2.1cm]{geometry}
\usepackage{caption}
\hypersetup{
    colorlinks=true,
    urlcolor=blue,
    linkcolor=blue,
    citecolor=blue
}

\title{Why Locking Up Cartel Members Does not Work}

\author{
  Rafael Prieto-Curiel\thanks{Complexity Science Hub, Metternichgasse 8, Vienna, Austria. \texttt{prieto-curiel@csh.ac.at}} 
}

\date{}

\begin{document}
\maketitle
%\linenumbers

\section*{Abstract}

One of the core strategies to reduce cartel violence is by directly targeting members with law enforcement. Whether targeting leaders, disrupting parts of the organisation, or incarcerating members, the purpose is to reduce the strength of cartels directly \cite{calderon2015beheading}. Most security strategies result in increased incarceration rates \cite{lessing2016inside}. Yet its effectiveness in addressing organised crime remains unclear, particularly if it fails to prevent recidivism upon release from jail \cite{maltz1984recidivism}. Here, a model is constructed to quantify cartel participation across generations, where individuals are recruited, age over time, and exit cartels as victims of a homicide or due to incapacitation, or retirement \cite{PrietoCampedelliHope2023}. Incarcerating cartel members prevents less than 10\% of cartel offences. Additionally, doubling penalties would reduce cartel members' potential by less than 5\%, thereby challenging proposals for stricter rules. Yet, rehabilitation after prison, often neglected as an integral part of the security strategy, could be more effective in lowering cartel crimes.

%%%%NOTES
%%%% FROM INEGI 2025 08 30
%%%% in 2020, of the 94777 males incarcerated, 18230 had a previous criminal record, so 19.23% of 
%%%% FROM ENPOL: Considera la probabilidad de regresar a prision: 4.5%

\section{Introduction}

%%%% incarceration
{
A central belief underlying incarceration is that it deters further offences by inducing fear of future punishment to the people incarcerated \cite{maltz1984recidivism}, or by exemplary punishment among other potential offenders \cite{d2015statistical}. However, its effectiveness is unclear, particularly related to gangs and cartels \cite{lessing2016inside}. Many groups leverage their power in prison to control life outside, including control over drug markets \cite{lessing2017counterproductive}, and have turned jails into centres of life and training for gang networks \cite{miguel2010central}. Thus, it is unknown whether experiences in jail and the stigmas associated after release reduce the likelihood of successful reintegration \cite{nagin2009imprisonment}. The limited effectiveness of sanctions reflects complex factors such as poor rehabilitation outcomes \cite{berenji2014recidivism}, perceived risks of apprehension \cite{tomlinson2016examination}, the deterrent function of penalties \cite{tsebelis1990penalty}, and group adaptation to incarceration \cite{helbing2015saving}. In recent years, the inmate population in Latin America has quadrupled, but contrary to expectations, this has not led to a decrease in criminality in the region \cite{Bergman2024PrisonsCrimeLA}. Incarceration often reinforces and transmits social inequalities across generations \cite{wakefield2010incarceration}, fosters criminal specialisation through contact with other offenders \cite{britt1996measurement}, imposes high budgetary costs \cite{schmitt2010high}, and has not proven effective in reducing violence in the region or curbing the trade of illicit drugs to other countries. 
}

%%% quantify the impact of incarceration
{
Since a core element of the strategy against cartels is to deploy armed forces \cite{blair2023little} with a subsequent incapacitation of its members, a critical question is how many cartel-related crimes, if any, are actually prevented through imprisonment \cite{MixedPoissonNagin}. Estimating the crimes averted during prison and the rate at which released offenders reoffend is essential to determine whether imprisonment has any measurable effect on cartel participation. If behaviours remain unchanged before and after incarceration, then the impact of this strategy is, at best, limited to the prevention of crimes during the period of imprisonment \cite{avi1973quantitative}. 
}

%%% birth cohort black box. A Birth cohort perspective is useful to determine the impact. Yet, for cartels is impossible. Black box
{
Age-cohort studies are a fundamental tool for understanding criminality. For example, by looking at criminal convictions in Glasgow during the 1950s \cite{YoungDelinquentFerguson}, or by following all males born in Philadelphia for more than 20 years, it was observed that most individuals rarely engage in crime and only a small group of the cohort is responsible for most offences \cite{ConcentrationWolfgangOneCohort, DucksWolvesEck}. Similarly, by matching 500 boys in correctional schools in Massachusetts with 500 non-delinquent boys, it was possible to study the causes and roots of criminal behaviour and to link them with family and social backgrounds \cite{glueck1950unraveling}. An age-cohort analysis of crime participation aims to detect many elements, such as who becomes a delinquent \cite{west1973becomes}, determine the age of criminal onset \cite{farrington1986age, lussier2012criminal}, and identify predictors of youth violence \cite{hawkins1998review}, such as experiences of abuse or neglect in childhood \cite{CycleOfViolenceScience}. In the case of mafias, for example, analysing thousands of offences and more than 10,000 Italian mafia members revealed that many individuals joined quite early in life, but co-offending increased with years of criminal experience \cite{meneghini2022co}. Similarly, a study of the prison population in Mexico showed a high frequency of domestic violence, often combined with excessive alcohol consumption, and incarcerated family members as triggers of crime participation \cite{Bergman2014DelitoCarcel}. However, it is nearly impossible to trace a birth cohort in the case of cartels. When a person is incarcerated or killed, it is rarely known whether they belonged to a cartel. Individuals who are incapacitated are typically convicted for a specific crime, such as murder, kidnapping, or extortion, and only seldom on charges of organised crime, which leaves their affiliation undocumented. 
}

%%%% age cohort and cartels
{
Here, we construct an individualised agent model of recruitment, participation, and incarceration for each cartel member over the past and upcoming decades, accounting for the flow of new and existing members. Each individual who enters a cartel is represented as an agent with specific attributes, including birth year and recruitment date. For each person, their age is sampled from a distribution that reflects a tendency toward early-onset delinquency. The model estimates the age distribution of active cartel members at any given time, as well as the ages of those who are recruited, killed, or incapacitated. Focusing on a specific age cohort in Mexico (males born in 1990) and using estimates of both the stock and the flow of cartel members, we estimate how many were recruited, how many will be killed or incapacitated, and the overall scale of cartel participation for that cohort. We quantify the cartel potential for that cohort, reflecting the total crimes committed during their affiliation with a group. Drawing on data from an extensive prison survey in Mexico, each incapacitated person is assigned a sentence length. From this, we estimate the high rate of reoffending among cartel members and assess that only 9\% of cartel potential is averted through incarceration. Finally, we compare the effectiveness of reducing cartel potential through longer prison sentences or social rehabilitation. Results show that focusing on better reintegration schemes proves far more effective for reducing the cartel potential than imposing harsher prison sentences. 
}

\section{Results}

% cartels must recruit g people per week to avoid collapsing. Assume that those recruits happen mostly among males and between 15 and 55 years (SM for the age and gender focus). Assume that the cartel career begins at 15 and ends by 55, so 40 years (SM for shorter and longer careers)
{
One of the key factors behind the rapid expansion of cartel activities in Mexico is their intense recruitment \cite{PrietoCampedelliHope2023}. A cartel may lose hundreds of members incapacitated by the state or killed by a rival group each week, but compensates for these losses by incorporating recruits at an ever faster rate \cite{PrietoCampedelliHope2023}. Recent estimates suggest that in 2022, Mexican cartels comprised 175,000 people and that during that year nearly 19,000 new members were recruited, while 6,000 were incapacitated and 7,000 were lost due to killings by rival cartels, disappearances, or retirements, resulting in a net annual increase of around 6,000 members \cite{PrietoCampedelliHope2023}. Mexican cartels experience a high turnover, as approximately 11\% of their membership is renewed each year. Building on the estimates that capture members entering (through recruitment) or leaving (as the victims of homicide or through incapacitation, or retirement), we approximate the size of cartels between 2005 and 2045 and how a specific age cohort participates with them. To analyse the participation of a birth cohort, a list of all active members (including their date of birth) is created, and a record is kept of each member's recruitment, death, incapacitation, or retirement. Each week, a number of new members, proportional to the combined size of the cartels, are recruited. Cartels tend to recruit mostly young males \cite{zavaleta2016aproximacion, SVyP2024TikTok}, so the age of each new member is sampled from a distribution that favours young individuals. Their age is assumed to fall between an entry age (taken to be $l = 15$ years old) and a late recruitment age (taken to be $u = 35$ years old). From the list of active members, some are killed (as a homogeneous process with probability $\kappa_k$), some are incapacitated (also as a homogeneous process with probability $\kappa_p$), and some retire (as an inhomogeneous process in which elderly members are more likely to retire). By constructing an individualised list of each recruited member, we estimate the share for every cohort (details in the Methods). Then, for each active member of cartels, we trace their age and follow them until they are either killed or they retire. Thus, instead of directly assigning ages to each active member, we assign birth dates, model their life courses, and obtain the age composition as a result (Figure \ref{Composition}). 
\begin{figure}[!htbp]
\centering
\includegraphics[width=0.5\textwidth]{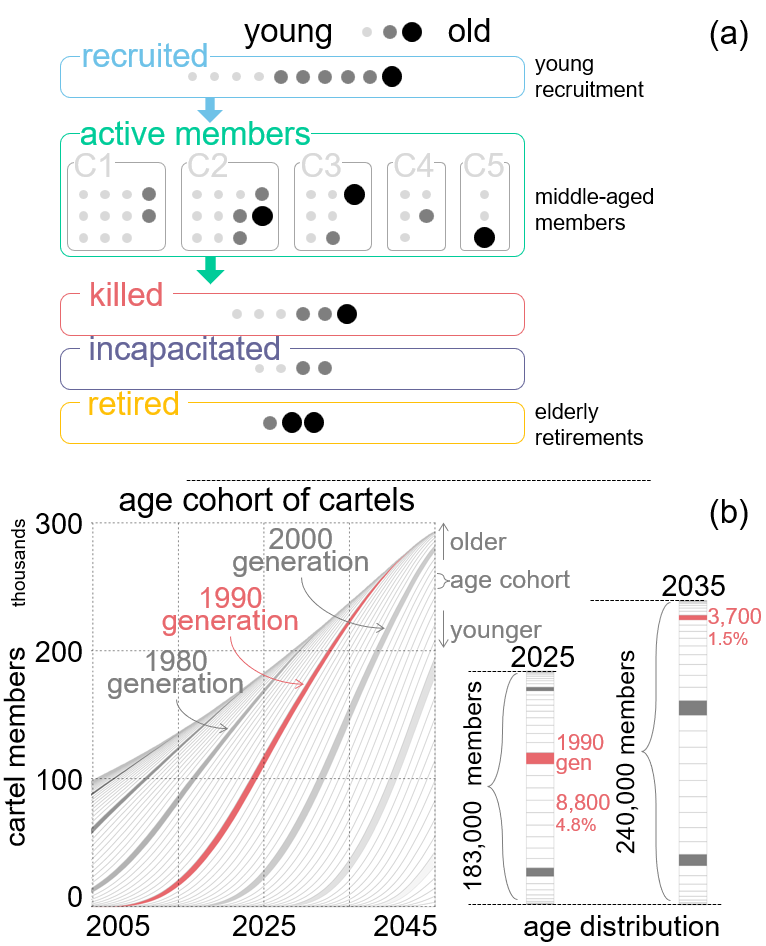}
\caption{Model for the flow of cartel members according to their age. a - Each week, a list of active members is updated by keeping track of members who are recruited (mostly young), who are killed, incapacitated, or retire. b - Composition of cartel members by age (vertical), between 2005 and 2045 (horizontal). Each stripe represents an age cohort (with the 1990 cohort highlighted). Younger cohorts are below older cohorts. We estimate that by 2025, 3\% of cartel members are under 20 years old, 38\% are between 20 and 30, 43\% are between 30 and 40, 15\% are between 40 and 50, and only 1\% are over 50 years old. Of the 183,000 active cartel members in 2025, 4.8\% are from the 1990 cohort. By 2035, active cartel members are expected to be approximately 240,000, including 1.5\% from the 1990 cohort. }\label{Composition}
\end{figure}
}

% focus on the 1990 cohort since they turned 15 in 2005 and they are halfway through their career in 2025. So for that cohort, the first recruitment occurred in 2006, when the war against cartels in Mexico started and with a sharp increase in homicides.
{
By examining the flow of cartel members, it is possible to estimate a cohort's engagement with organised crime. We focus on the male cohort from 1990, since they were among the first to be exposed, as teenagers, to the war against organised crime (launched in Mexico in 2006). Due to a decline in fertility (from 3.7 to 2.3 births per woman between the 1960s and 1990s), this group is also among the largest cohorts in the country, with approximately 1.35 million males (compared to 0.9 million males born in 2024) \cite{UNPopulationData}. Additionally, assuming that late cartel recruitment occurs for individuals who are $u = 35$ years old, the 1990 age cohort has now completed its recruitment stage, allowing us to estimate how many of those males were recruited. We estimate that in 2005, approximately 11,000 new members were recruited by cartels, of whom the 1990 age cohort (only 15 years old) contributed only 0.2\%. However, as years passed, cartels recruited more members, and that cohort contributed to a larger share. In 2015, cartels recruited over 15,200 members, of whom the 1990 cohort accounted for 7.5\% (its peak). By 2025, cartels had recruited approximately 20,500 members; however, the 1990 cohort was already too old for cartel recruitment, contributing only 0.2\% to the total (Figure \ref{Composition}). 
}

%%% total for G1990
%%% add a citation that UN Life expectancy
{
Results show that between 15,000 and 16,000 people from the 1990 cohort will have participated in cartel activities at some point in their lives, corresponding to 1.1\% of the males from that group (Figure \ref{Gen1990Trajectory}). By tracing the cartel career since recruitment until termination, we find that of all individuals that a cartel recruits, 41\% will be killed, 34\% will be incapacitated, and 25\% will retire without spending time in prison. The cartel career is highly lethal and ends the lives of its members quite early, since they are killed on average at 33.6 years old. The life expectancy of the males recruited by a cartel decreases by 14.4 years compared with those who were not recruited, from 69 to 54.6 years (Figure \ref{Gen1990Trajectory}). 
\begin{figure}[!htbp]
\centering
\includegraphics[width=0.5\textwidth]{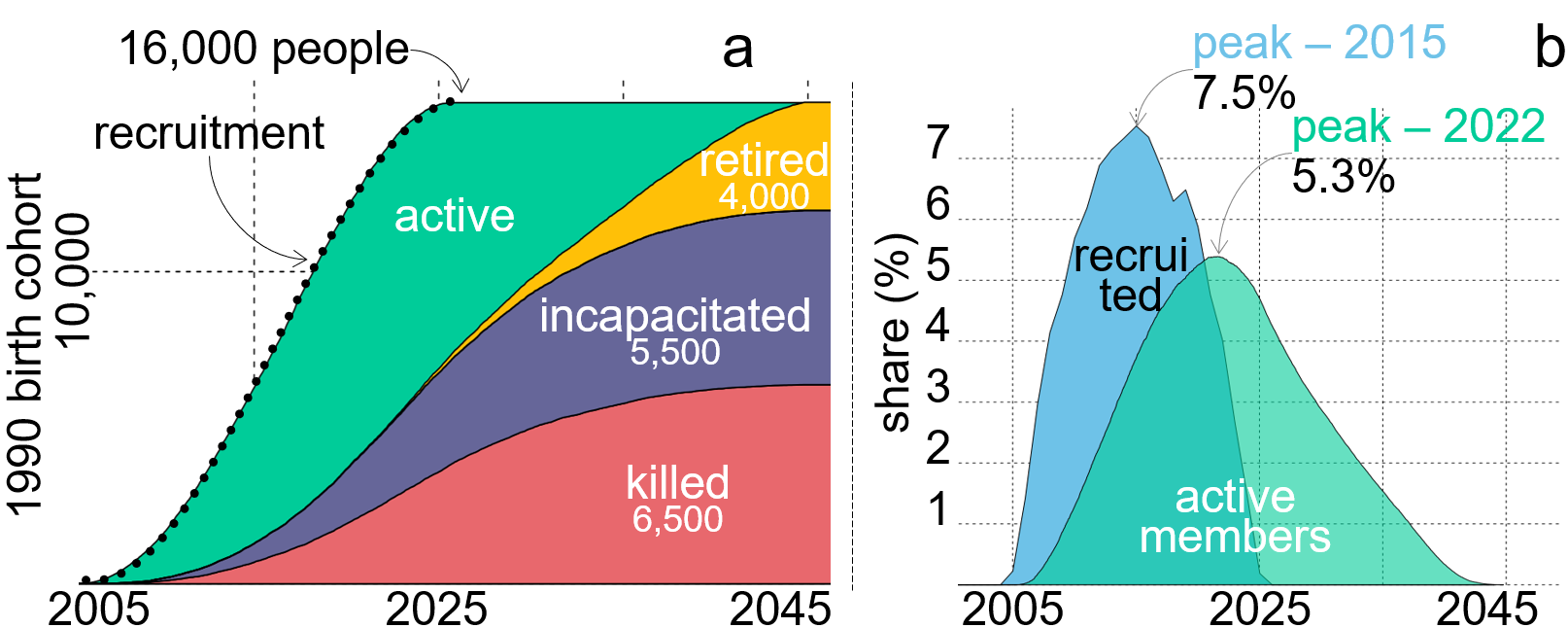}
\caption{Life course trajectory of the 1990 cohort. a - Share of the different states of the recruited individuals of the 1990 age cohort between 2005 (when that cohort experienced its first recruitment) and 2045 (when all individuals have exited cartels, either being killed, incapacitated, or retiring). b - Share of the recruits and the active members that belong to the 1990 cohort between 2005 and 2045.}\label{Gen1990Trajectory}
\end{figure}
}

% the 1990 cohort contributes to f% of the active cartel members in 2025. In the past 20 years, they have suffered x% of cartel homicides.
{
Although recruitment of the 1990 cohort is nearly complete by 2025, it is one of the largest cohorts among active members that year. That age cohort reached its peak level of active cartel members in 2022 (5.3\% of active members) and has since decreased. By 2025, there are 8,800 active cartel members from the 1990 cohort, corresponding to 4.8\% of the total size (Figure \ref{Gen1990Trajectory}). By 2035, only 3,700 people from that cohort will remain active (1.5\% of the cohort), and by 2045, most of them will have completed their careers. 
}

%%% validate, black box. 
{
Validating the results with observed data is challenging for several reasons. Victims of a homicide are not labelled based on whether they belong to a cartel or not. And a similar thing happens with incapacitations. Except for rare cases (usually highly mediatic ones, such as the head of a cartel), most remain silent, and their cartel affiliation is undisclosed in official records. However, assuming that a share of the homicides in the country are members of a cartel and they have the same age distributions, we can take the observed and the modelled age of the victims of homicide and determine how similar they are. Reported data \cite{InegiMortalidad} for ten age groups (groups of people between 15 and 19 years old, 20 and 24, and so on, until the group of people between 60 and 64 years old, as reported by official statistics) between 1990 and 2024 shows a high correlation with the age distribution of the modelled homicides (details in the Methods). Similarly, data for nine age groups (people between 18 and 24 years old, 25 and 29, and so on, until the group of people above 60 years old, as reported by official statistics) shows a high correlation with the age distribution of people who are incarcerated each year between 2017 and 2024 \cite{CNSIPEE}. Although creating a list of recruited members depends on assumptions about the age distribution for recruitment and the dynamics of homicides, incapacitation, and retirement, the results show only a slight variation across different distributions (details in the Methods). 
}

\subsection{The limited benefits of incarceration}

% suppose that a cartel members spends x years in prison and goes back to the cartel after those years (so no impact on reducing membership) and assume that incapacitation happens at random among the cartel members. Then, of the CC people who will spend time in prison due to cartel activities, cc will be incarcerated more than once. Maybe sample from enpol?
%%% take this to the SI: in the ``Ley Federal Contra la Delincuencia Organizada'' (Federal Law Against Organised Crime) 
{
To quantify the impact of incarceration on cartel crime, the individuals who are imprisoned and the dates of incarceration are considered. The Mexican Federal Penal Code defines organised crime when three or more persons organise to commit crimes like terrorism, drug-trafficking, illicit arms trafficking, and others \cite{codigoPenalMexicano}. The penalty for organised crime ranges from 4 to 40 years, depending on the severity of the crime and whether the person held leadership roles \cite{LFCDO}. In reality, it is often impossible to sentence a cartel member due to organised crime, so less than 3\% of the incarcerated population receives that sentence \cite{CNSIPEE}. If cartel members are processed, they may be sentenced due to other felonies besides organised crime (such as homicide or kidnapping). Additionally, it is not known how frequently a judge determines whether a person committed a serious crime, whether they had a leadership role, and how often that sentence is combined with other types of crime (so it is not known the frequency with which sentences are assigned to cartel members). To reproduce the process of incarceration as realistically as possible, we use data from the largest survey of the prison population in the country, which includes reported crime type and sentence length \cite{ENPOL}. To each case, we randomly assign one of the observed sentences and the time spent in prison. The principle is to capture real sentences from incarcerated people and their variability (details in the Methods). We consider the scenario in which a member of a cartel is convicted and imprisoned for the length of their sentence, and then they can either die in prison (if the age of the person is combined with a long sentence) or they are released (and then model their lives after).
}

%%% results from reincap. 
%%%% write somewhere in the SI $9.7 \pm 6.9$ years
{
Approximately one in three cartel members will be incapacitated. On average, it takes 9.7 years from their recruitment to their incapacitation. Then, incarceration lasts an average of 8.7 years. Based on the rate and length of sentence, we estimate that among all age cohorts, prisons have approximately 50,000 cartel members (of a current prison population of 221,000 males). 
}

{
Prisons in Mexico are severely understaffed \cite{insightcrime_closing_prisons_mexico} and suffer from overpopulation and violence among inmates \cite{cndh_dnsdp2019}. Some cartel members are capable of operating effectively from prison, maintaining transnational ties and street presence \cite{miguel2010central}. Thus, they have a limited impact, if any, regarding social rehabilitation upon release \cite{lessing2017counterproductive}. To quantify the best-case scenario for the effect of incarceration, we assume that cartel members are inactive while in prison and rejoin the group upon release. After rejoining, they follow the same dynamic as the other members, meaning that some of them will be killed or incapacitated with the same probability as the other active members. 
}

%%%% results for the incarcerated people
{
Results show that, upon release, approximately 25\% of individuals who have been imprisoned once and released will be reincarcerated. This relatively low rate is not because of any rehabilitation (assumed to be negligible given the current conditions) but because most cartel members avoid incapacitation. The estimate is similar to the share of individuals with a criminal record who are convicted \cite{CNSIPEE}, as approximately 24.5\% of imprisoned males in 2024 had a prior criminal record (see the SI for details). Spending several years in prison reduces the chances that they will be killed (from 41\% to 31\% after incarceration), and increases the likelihood of retirement (from 25\% to 41\%). Approximately 4\% of them will die in prison due to their age and the length of their sentence (Figure \ref{LifeAfterPrison}). Based on the incarceration and reincarceration estimates, we calculate that cartel members from the 1990 cohort will spend 60,300 years in prison (an average of 3.8 years in jail for all members).

\begin{figure}[!htbp]
\centering
\includegraphics[width=0.5\textwidth]{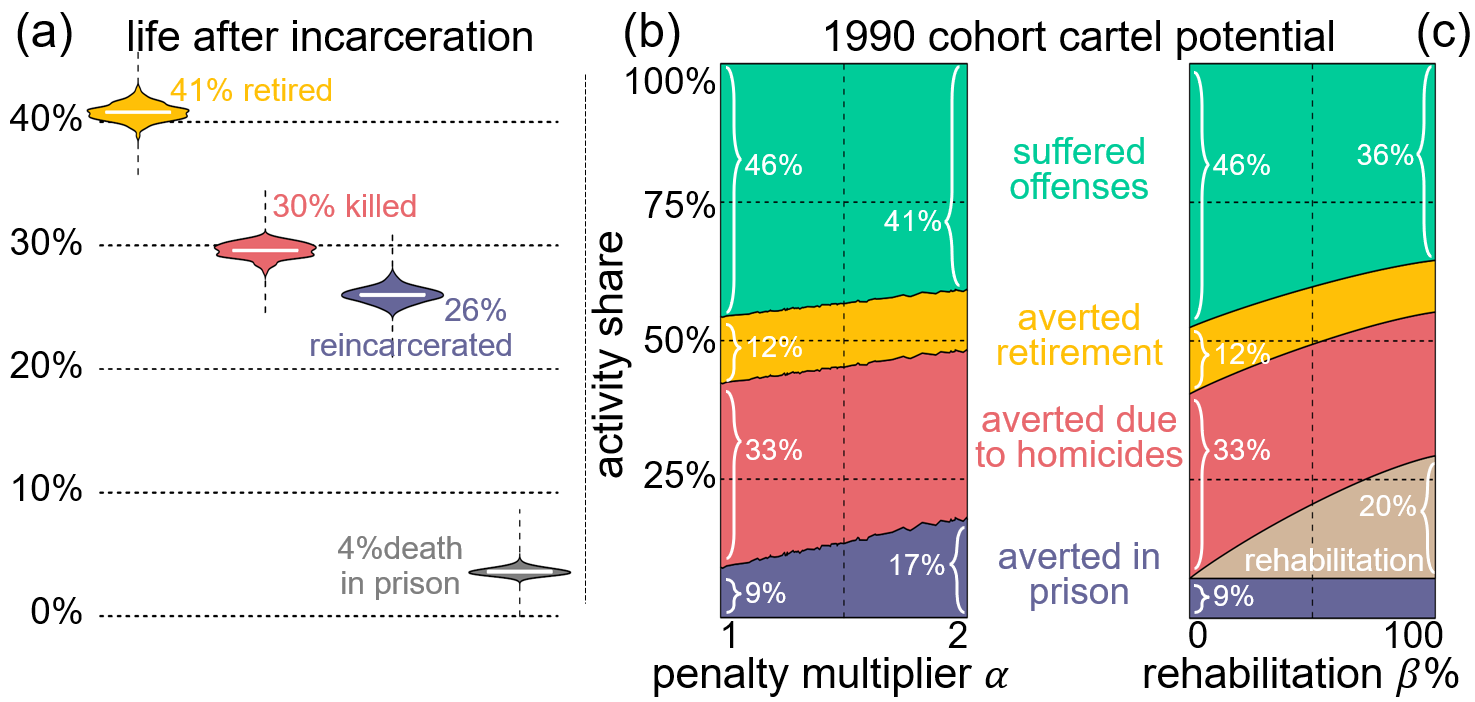}
\caption{Modelled life after prison. a - Distribution of ending states after spending time in prison for all individuals from the 1990 age cohort who were incapacitated. Intervals are obtained by repeating the assignment of sentences to different cases. b - Impact of increasing incarceration penalties (horizontal axis) on the cartel potential of the 1990 age cohort (vertical). By multiplying each sentence by a penalty multiplier $\alpha\geq 1$, the cartel activities of the cohort vary. With the current penalties (with a value of the multiplier $\alpha = 1$), of the cartel potential of the cohort, 33\% is averted due to homicides, 12\% due to retirement, and only 9\% is averted due to incarceration (so 46\% of the years of cartel activity will be effective). By doubling the incarceration rate (with $\alpha = 2$), the prison prevents 17\% of the cohort activities, reducing the share of activities averted by retirement and homicides, so the effective cartel offences drop from 46\% to 41\%. c - Impact of increasing social rehabilitation (horizontal axis) on the cartel potential of the 1990 age cohort (vertical). With $\beta = 0$, all the prison population rejoins the cartel upon release, and with $\beta = 1$, no one returns to the cartel, reducing its potential from 46\% to 36\%. }\label{LifeAfterPrison}
\end{figure}
}

% of the 1.4 million people, Incarceration will be for hh%
%%%% severity and intensity
{
To quantify the impact of incarceration on cartel crime, we proxy for the adverse effects of each cartel member by their active time in the group (details in the SI). We define cartel potential as the total time a cohort could spend in the organisation from recruitment until natural death. Given their recruitment date and life expectancy, a cartel member could participate in its activities for approximately 42 years. Thus, the 1990 age cohort has a total cartel potential of 670,000 years. Of this potential, 33\% is averted because its members are killed, 12\% because its members retire, and only 9\% is averted because its members are incarcerated (details in the SI). Thus, only 46\% of the potential of the cohort is suffered in the country (Figure \ref{LifeAfterPrison}). Due to a combination of low incapacitation rates, short sentences, and the lack of social reintegration after prison, incarceration (one of the core security strategies in Mexico) reduces, at most, 9\% of the cartel potential of the 1990 cohort. 
}

%%%% increasing sentences and populism
{
Increasing penalties is frequently proposed as a strategy to avert more cartel crimes \cite{bonner2019tough}. Standard narratives (often referred to as Mano Dura \cite{cutrona2024conceptualizing}) include more incapacitations \cite{VilaltaFondevila2019}, longer prison sentences \cite{huhn2017punitive}, and even extreme measures, such as life sentences or the reintroduction of the death penalty \cite{LuciaDammertBuekele}. To examine the impact of rising penalties, we assign a sentence to each incapacitated person and multiply the length by a factor, $\alpha \geq 1$. We assume that longer sentences have no impact on the recruitment capacity of a cartel or on life after prison, so the effect is only observed during the time spent in jail (Figure \ref{LifeAfterPrison}). Results show that doubling the length of assigned sentences would decrease the cartel potential of the cohort from 46\% to 41 (details in the SI). Increasing sentence length would only marginally reduce cartel activity, but would worsen prison conditions and could help prison gangs establish authority outside prison and organise criminal markets \cite{lessing2016inside}. 
}

%%%% rehab. but check 36% or 29
{
A critical element in a country's security is social rehabilitation. Many jails in Mexico are an overpopulated epicentre for gang training and retaliation between rival groups, whilst maintaining parts of their power on the streets \cite{lessing2017counterproductive}. Under current conditions, social rehabilitation is negligible, and therefore, it is assumed that all cartel members released from prison will rejoin the group. However, if some individuals were rehabilitated, the outcomes could differ significantly \cite{butorac2017challenges}. By designing prison-based rehabilitation and reintegration strategies delivered upon release, as well as community-based programmes, the outcomes of prison could be more effective \cite{massa2025works}. We quantify the impact of social rehabilitation by considering how cartel potential decreases when a share, $ \beta$, of individuals are reintegrated into society and stop contributing to it. With $\beta = 0$, no rehabilitation occurs, whereas with $\beta = 1$, nobody released from jail returns to the cartel (Figure \ref{LifeAfterPrison}). Currently, 9\% of the cartel potential is prevented by incarceration, but it could expand to 29\% if all members are rehabilitated. With full rehabilitation ($\beta = 1$), the cartel potential suffered in the country would drop from 46\% to 36\%, making it far more effective than doubling prison sentences.
}

\section{Discussion}

% oc is devastating
{
Organised crime cartels have had devastating consequences globally. They have become the leading traders of illicit drugs globally, and with the flow of unregulated substances, drug overdoses have become one of the top priorities worldwide. In the United States, their largest market, drug overdoses claim approximately 100,000 lives annually, a figure that has doubled within a decade \cite{klobucista2023fentanyl}. What is experienced primarily as the circulation of substances, the control of black markets, and a deadly overdose pandemic on one side of the world has turned into a fierce conflict between groups on the other. Cartels have contributed to making Mexico one of the most violent countries, recording one of the highest murder rates globally (approximately ten times higher than the European rate \cite{UNODC}). Cartels in Mexico have killed thousands and displaced millions of people. The country recorded more than 30,000 homicides in 2024, and nearly 20,000 people went missing that year and remain unaccounted for. The number of people who fall victim to homicide or disappear has almost doubled over the past decade. A substantial part of this violence stems from fierce conflicts between rival cartels. These groups fight for control of strategic areas used for the production, transformation, and transport of illicit drugs, but they also kidnap and extort people in exchange for protection \cite{santos_cid2025rehenes}. 
}

% yet, the demographic impact is unknown. 
{
Cartels are not static structures but highly dynamic organisations, with a constant flow of new members being recruited and former members leaving \cite{PrietoCampedelliHope2023}. The relative size of cartels today is estimated at approximately 0.26\% of the country's male population. Yet this does not mean that such a small fraction of the population is the only group engaged with a cartel. This effect arises not only because cartels are expanding in size, but mainly because they sustain rapid turnover among their members. We modelled the flow of members, tracing each person since recruitment until they are killed or retire. Approximately 1.2\% of males of the 1990 cohort will participate in a cartel.
}

%% cartels and facts about the career
%% demographic
{
The demographic cost of cartels is devastating. Involvement with a cartel reduces life expectancy by over 14.4 years due to homicide, while members spend an estimated 3.8 years incarcerated. The 1990 male cohort will experience a lethality exceeding 600 homicides per 100,000 people attributable to cartel involvement. Among all recruited people, half will be killed (80\% without prior imprisonment and 20\% after a period of incarceration). Only one in four individuals retires without being killed or incapacitated, and an additional 14\% eventually retire following imprisonment. 
}

%%%% measure cartel activities
{
Given the secretive nature of cartel operations, the constantly changing conditions, and the difficulties in identifying casualties or incapacitations among members, the results should be interpreted as model-based approximations of a highly complex reality. Nevertheless, the observed age distributions of those killed and those who are incarcerated show a strong correlation with the ages predicted by the model. Assuming that incapacitated cartel members are imprisoned for a period and then return to the organisation upon release, the model replicates the observed reincarceration rate. In the model, we assume that all members return to the cartel after serving their sentences (so no rehabilitation), and, given the high turnover of cartel members and low incarceration rates, only about one in four incarcerated cartel members has a prior criminal record. This finding underscores the importance of correctly interpreting the proportion of incarcerations with criminal records, and the estimate should be used as a benchmark for assessing whether incarceration has any rehabilitative effect. 
}

%%% cartel potential and prison
{
The age-cohort model enables us to analyse the duration of membership of all its members, referred to as cartel potential. The metric serves as a proxy for estimating the damage caused by those members. The 1990 age cohort has an estimated cartel potential of about 670,000 years. Of the cartel potential, roughly one-third is averted because members kill one another (the most significant factor reducing cartel potential), followed by early retirement and, to a lesser extent, incapacitation. One of the core elements of the security strategy is to incapacitate cartel members; however, this approach affects only a small portion of their operations, as few members are imprisoned, they spend relatively little time in custody, some even continue their activities while incarcerated, and then return to the cartel after release. 
}

%%%% rehab
{
Would longer prison sentences reduce cartel activities? We estimate that doubling prison sentences, while keeping recruitment rates and other parameters constant, would reduce cartel potential by less than 5\%. In practice, prisons do little to deter reoffending or to prevent new recruitment through exemplary punishment. Longer sentences could exacerbate the overcrowding in the country's already saturated prisons. In the US, for example, stricter sentences for drug-related offences are one of the main reasons behind skyrocketing incarceration rates \cite{schmitt2010high}. Also, individuals with longer prison sentences experience greater deterioration of their social relationships and encounter more difficulties when they return to the community, increasing the chances of reoffending \cite{UNODC2018}. Correctional institutions are often regarded as enforcers of punishment rather than integral parts of the social reintegration process \cite{tajuddin2025forgotten}. Incarceration frequently has serious collateral effects: offenders may lose their jobs, possessions, housing, health, social ties, and even develop mental health issues or self-defeating habits \cite{UNODC2018}. Without rehabilitation, recruited people are often caught up in a vicious cycle of violence, incarceration, poor social integration, reoffending, and reconviction \cite{UNODC2018}. But also, without rehabilitation, longer sentences yield only marginal reductions in violence \cite{becker1968crime15000citations, d2015statistical}. Rehabilitation is a crucial component of a country's security strategy, which is often overlooked \cite{butorac2017challenges}. By carefully designing mechanisms to help past offenders return to the community (including pre- and post-release monitoring), incarceration and reintegration could be a powerful tool to prevent offences \cite{UNODC2018}. Incarceration without rehabilitation only averts 9\% of the cartel potential of the cohort, but that number could triple if individuals were rehabilitated and reintroduced into society after prison. 
}

\clearpage

\section{Methods}

\subsection{Changes in cartel size}

%%% intro to the method with a two-step process
{
To estimate the cartel participation of some age cohort $G_x$ at some time $t$, two sequential steps are followed. First, five time series are produced, which are those recruited $R(t)$ each week, the aggregate number of active members $A(t)$, and then those who are incapacitated $I(t)$, retired $S(t)$, and killed $K(t)$. Then, based on the aggregate estimate (indicating the number of individuals recruited on a specific date), five lists are created and maintained by sequentially assigning a date of birth to each recruited person and then updating them according to their stage in the dynamics. The process of creating an active list of members and updating that list each week until 2045, based on how many people are recruited, and then killed, incapacitated, or retired, requires creating and tracing nearly one million agents, keeping track of the date of birth of each, as well as the date at which they entered and exited a cartel. Each step is detailed below.
}

%% grouping all cartels, we obtain the expression
{
By looking at open sources, including national and local newspapers and narco blogs, over 150 active cartels in Mexico were identified in 2020 \cite{BacrimMexico}. The interactions between cartels are complex and include alliances, rivalries, and fragmentation, which often spark deadly conflicts among them \cite{prietoDieter2025reducing}. Considering the combined number of members across all cartels and the reasons for these changes (recruitment, homicides, incapacitations, and retirements), it is possible to model the combined cartel size as a (single) indicator that changes over time. The changes in cartel size can be expressed as
\begin{equation}\label{eq:MasterEquation}
\dot{A} (t) = \underbrace{\rho A (t)}_\text{recruitment} - \underbrace{\eta A(t)}_\text{incapacitation} - \underbrace{\theta A (t) }_\text{conflict} - \underbrace{\omega A^2 (t)}_\text{retirement},
\end{equation}
where $\rho$ is the recruitment rate of cartels, $\eta$ is the incapacitation rate, $\theta$ is the rate at which members kill each other, and $\omega$ is a retirement (or saturation) term, which captures members who leave a cartel for reasons besides incapacitation and homicide. A similar system was proposed that considers all identified active cartels and their interactions \cite{PrietoCampedelliHope2023}. Since the interest here is in capturing a cohort that participates in cartel activities (regardless of which cartel), the interactions are assumed to collectively follow a pattern captured by the parameter $\theta$. In Equation \ref{eq:MasterEquation}, the term $R(t) = \rho A(t)$ gives the time series of the people who are recruited each week, the term $I(t)=\eta A(t)$ gives the series of incapacitations, the term $K(t) =\theta A(t)$, and $S(t) = \omega A^2(t)$ gives the number of people who retire. 
}

% we simulate weeks of cartel dynamics
{
Based on the observed number and the increasing trends of missing people \cite{RNPDNO}, homicides, and incapacitations in Mexico between 2012 and 2023 \cite{InegiMortalidad}, it was estimated that cartels had approximately 114,000 members in 2012 and 175,000 members a decade later \cite{PrietoCampedelliHope2023}. Additionally, it was estimated that cartels recruited 19,300 new members in 2023 and experienced about 6,000 incapacitations and 7,000 losses due to inter-cartel killings. We use those estimates to approximate values of the parameters $\{ \rho, \eta, \theta, \omega \}$. The values used reproduce, as closely as possible, the observed trends of homicides, missing people, and incapacitations in Mexico. Although many fluctuations and shocks may rapidly alter the cartel system in the country \cite{prietoDieter2025reducing}, we simplify these complex interactions by assuming a global trend across all cartels (details on cartel dynamics are provided in the SI).
}

\subsection{Generational distribution of cartel dynamics}

% how to distribute recruitment among different age groups 
%%% FIGURE WITH PHI AND ASSIGNMENT
{
Having determined the time series for recruits $R$, active members $A$, incapacitations $I$, retired $S$, and killed $K$, a list of all active members is created by assigning a precise date of birth to each agent since the day they were recruited until the day they exit the system (either natural death or being killed). The time series $R(t)$ represents the number of people recruited during week $t$, so those agents are added to the list of active members. Their ages are sampled from a distribution that assigns each person an age $a$ in years. Each week, the age of each person recruited is sampled from a distribution $\phi(a)$. We take the distribution $\phi(a)$ as 
\begin{equation} \label{AgeFunctionMethods}
\phi(a) = s ( a -l) (u -a) I_{[l,u]}(a),
\end{equation}
where $l$ is the lower age for cartel entry and $u$ is the upper age for recruitment (i.e., the oldest age members are recruited). The function $I_{[l,u]}(a) =1$ for values of $a$ inside the $[l, u]$ interval, and zero elsewhere. The value of $s = 1/ \int_0^{100} ( a -l) (u -a) I_{[l, u]}(a) \, da= 6/(u - l)^3,$ guarantees that $\int_0^{100} \phi(a) \, da =1$, so it forms a probability distribution (Figure \ref{AgePyramidCartels}). Here, we set $l = 15$ as the entry age and take $u = 35$ for late recruitment, assuming that if a person is not recruited by age 35, they will never be recruited. A sensitivity test for the age of entry and recruitment termination is part of the SI. With the sampled age $a_i$ for each recruited member $i$, their date of birth is computed as $b_i = t - a_i$. Each member's date of birth is then recorded and used for the rest of the process.

\begin{figure}[!htbp]
\centering
\includegraphics[width=0.5\textwidth]{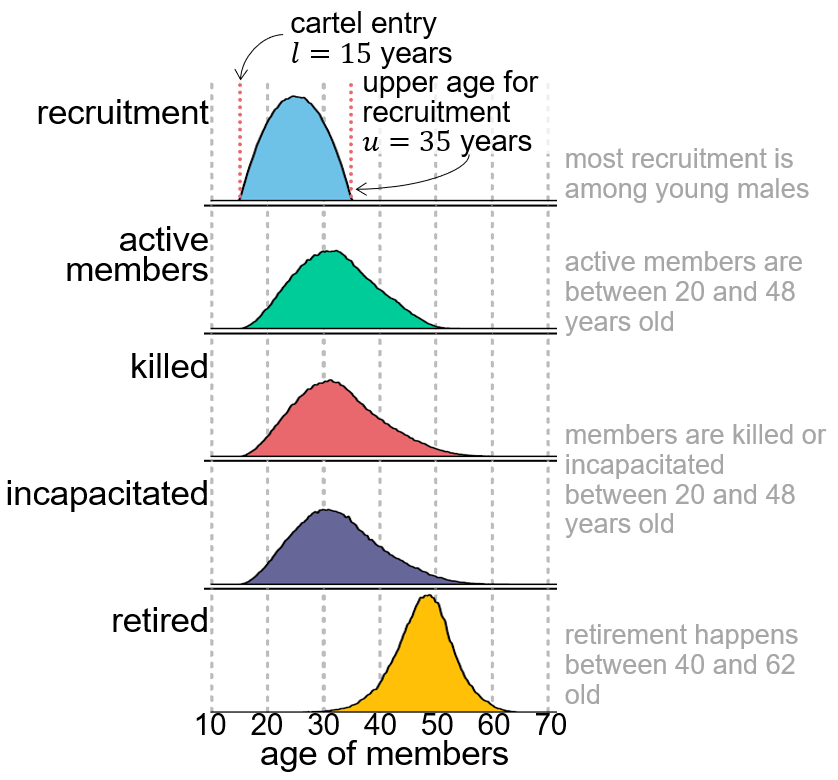}
\caption{Age distribution for the age of recruitment $\phi(a)$. The function $\phi(a)$ assigns an age to each recruited person between $l$ and $u$ years old. The dynamics then give a distribution of the age of active members, individuals being killed, incapacitated, and retired. Most recruits are young, whereas retirement typically occurs among elderly cartel members. }\label{AgePyramidCartels}
\end{figure}
}

%%% update the list
{
To update the list with all active cartel members, it is expanded by adding $R(t)$ members, each with their corresponding birthdate, $b_i$. Then, some members are removed from the list of active members, first by randomly sampling $K(t)$ individuals and moving them to the list of killed members. Similarly, $I(t)$ individuals are randomly sampled and moved to the list of incapacitations. To update the list of active members due to retirement, $S(t)$ are sampled and moved to the list of retired members, however, instead of randomly taking individuals, we take that the probability that the person $i$ retires is proportional to $a_i^\mu$ for some value of $\mu > 0$, so older individuals are more likely to retire (details in the SI). To the lists of recruited, killed, incapacitated, and retired, the date of the event ($t_0$) is registered. Each week, the age of active members is updated by adding one extra week of life. 
}

%% the results are
{
The assignment results in five lists. One list contains all recruited members, including the date they joined the cartel. One list has all the killed people, including the date of birth and date of their homicide, and similarly for incapacitations, retired people, and those members who are still active. Although the focus is on cartel dynamics between 2000 and 2045, the period is extended by two decades to obtain the age distribution of active cartel members in 2000. The age distribution in the year 2000 is then used to update the lists of active members. In total, the list of members who were recruited by a cartel between 2000 and 2045 includes 760,000 people. As the number of active cartel members increases, younger age cohorts are participating in greater numbers (Figure \ref{ShareByYear}).

\begin{figure}[!htbp]
\centering
\includegraphics[width=0.4\textwidth]{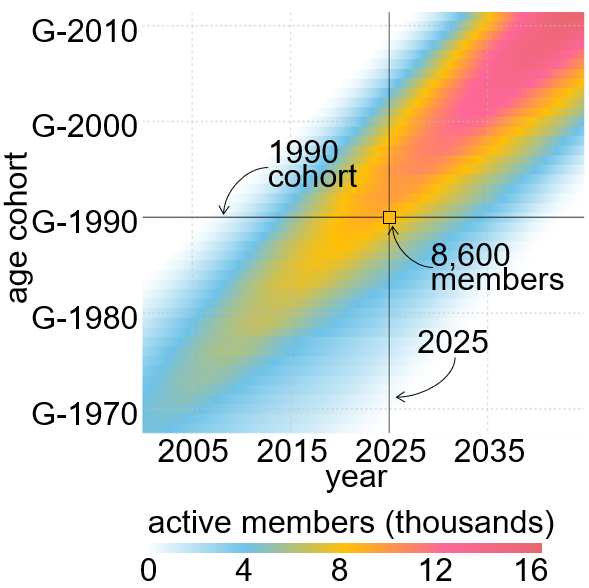}
\caption{Number of active cartel members for each cohort (vertical axis) between 2000 and 2045 (horizontal axis). In 2025, the 1990 age cohort had 8,600 cartel members, which decreased in the following years due to a low recruitment of that cohort and exits through retirement, incapacitation, or homicide. }\label{ShareByYear}
\end{figure}
}

\subsection{Age of homicides, incarceration, and retirement}

%%% other expressions for the cohort distribution (SM)
{
The age of recruitment is assumed to follow the distribution described by Equation \ref{AgeFunctionMethods}. Then, the list of active members is updated, and the age of all individuals is increased by one week. The members who are killed or incapacitated are randomly picked from the list of all members, so their age distribution is the same, but we assume that older cartel members are more likely to retire. Formally, for a member with age $a_i$, the probability that he is selected during that week to retire is proportional to $a_i^\mu$ for some value of $\mu>0$ (details in the SI). The processes of recruitment and retirement alter the age distribution of active members (by introducing younger members and retiring older ones). 
}

{
The obtained age distributions for the people who are killed and for the incapacitations can be compared with the official reports from the country. Taking the data for intentional homicides and for the people who joined the penitentiary system each year, we compare how similar both distributions are. In the case of the data for intentional homicides, the official records are available between 1990 and 2024 and divide the victims into ten age groups (including people between 15 and 19 years old, 20 and 24, and so on, until the group of people between 60 and 64). Thus, the data correspond to 34 years of observations, with 10 age groups per year (a total of 340 observations). Since the data for intentional homicides also includes many non-cartel-related incidents, the focus is not directly on the volume, but rather on verifying whether the observed age distribution \cite{InegiMortalidad} is similar to what our model predicts. To quantify the overlap between the observed numbers (say $x_{g,t}$ for some age group $g$ during year $t$) and the modelled numbers (say $m_{g,t}$), we adjust the model $m = \psi x$, for some value of $\psi$ (a linear model with no intercept). Approximately 27\% of the observed homicides were captured by the model, and the distributions are strongly correlated (Figure \ref{ObservedModelled}). 

\begin{figure}[!htbp]
\centering
\includegraphics[width=0.5\textwidth]{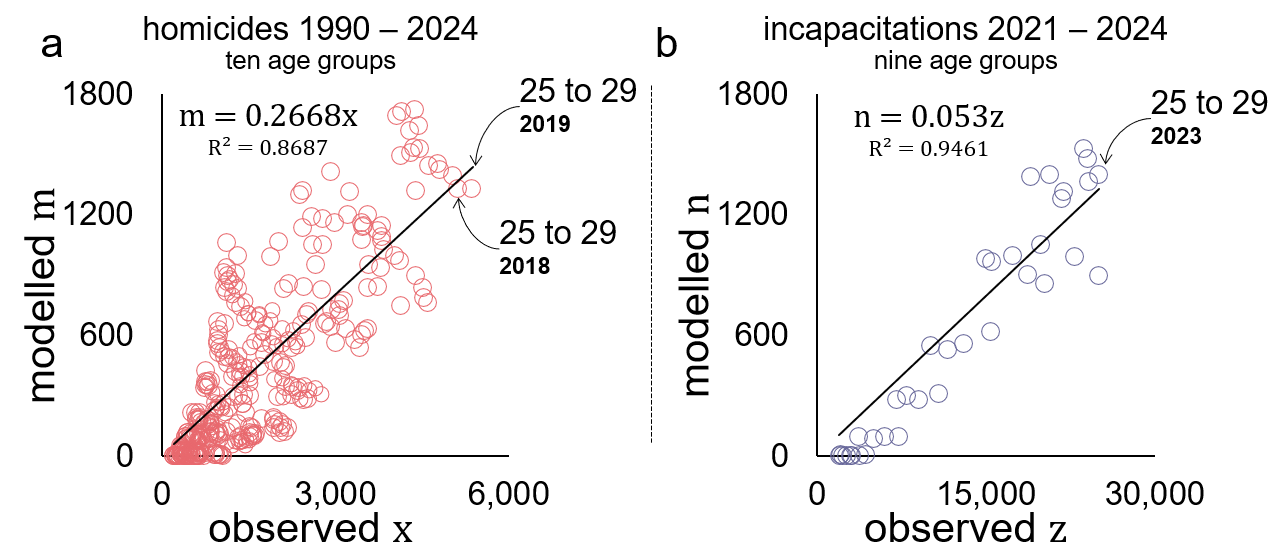}
\caption{Left - Observed number of homicides for each year and age group between 1990 and 2024 \cite{InegiMortalidad} (horizontal) and modelled number of homicides (vertical). Right - Observed number of incapacitations for each year and age group between 2017 and 2024 (horizontal) and modelled number of incapacitations (vertical).}\label{ObservedModelled}
\end{figure}

We follow a similar procedure with all people who are incapacitated each year. Data for the Penitentiary system groups incarcerations into groups of people between 18 and 24 years old, 25 and 29, and so on, until the group of people above 60 years old \cite{CNSIPEE}. Data is available from 2017 to 2024, covering 8 years and nine groups each (for a total of 72 data points). We adjust the model $n = \iota z$ for some value of $\iota$ (so also a model with no intercept). Approximately 5\% of the observed incapacitations captured by the model, along with a similar age distribution (Figure \ref{ObservedModelled}). 
}

\subsection{Time spent in prison and reoffending}

%%%% life in prison intro
{
To estimate how much crime is avoided in prison, it is necessary to determine many elements regarding cartel participation (Figure \ref{CartelCycle}). These elements include the frequency with which a person commits crimes (intensity), the impunity rate, the length of the sentence, and the reincidence rate \cite{avi1973quantitative}. Considering the number of crimes over a given period as a distribution with homogeneous intensity, and the type and severity of each crime as a process involving mutually exclusive outcome categories occurring over repeated trials, we can model the crimes averted in prison as a function that depends only on the length of the sentence (details in the SI).

\begin{figure}[!htbp]
\centering
\includegraphics[width=0.5\textwidth]{figures/CartelCycle}
\caption{Cartel cycle from crime participation to incarceration and reoffending. The model relates the number of crimes for each cartel member to the time spent as an active member. With no rehabilitation, only some crimes are averted during prison time.}\label{CartelCycle}
\end{figure}
}

% how to simplify from those who were already in prison? Enpol and time
{
To model the effect of prison for cartel members, it is necessary to assign the length of the sentence for each case. Using the largest survey of the prison population in the country allows us to capture the types of crime and the lengths of sentences \cite{ENPOL}. When a person is incapacitated, it is often impossible to determine their affiliation with a cartel. However, some of the members might be sentenced due to a crime they committed (for example, intentional homicide) without being able to determine their affiliation with a cartel. Thus, the length of the sentences must be modelled rather than inferred directly from data.
}

%%% by type of crime
{
The time spent in prison varies considerably depending on the type of crime for which the person is convicted, ranging from only a few months for minor offences, 10 years for robbery or shoplifting, 16 years for car theft, to nearly 30 years for crimes like kidnapping, extortion, organised crime, and intentional homicide. When examining a survey of individuals in prison, those with longer sentences are more likely to have been surveyed than those with shorter sentences (details in the SI). %To illustrate this, consider a scenario in which the sentences are either 1 year or 20 years in prison. If we observe 50 people with a 20-year sentence and 50 people with a 1-year sentence, it does not follow that half of the sentences are 20 years. We need to estimate the frequency with which sentences are enforced. In the example of one- and 20-year sentences, equal frequency means that one-year sentences are 20 times as likely to be assigned as the longer ones (since they are less likely to be monitored through surveys). 
}

%%%% to each case and repeat
{
We first estimate the frequency with which different sentences are assigned. Then, for each incarcerated person, one sentence is randomly selected and used as the length of their prison stay (ignoring sentence reductions for good behaviour or escapes). We then consider what may occur after incarceration. To compute the frequency of events after prison, we analyse the distinct scenarios. The first scenario happens when the length of the condemn and the age of the person exceed the life expectancy of that cohort (estimated at 69 years \cite{UNPopulationData}). This happens, for example, for people who are incarcerated at age 50 and are assigned over 20 years of prison. That scenario is labelled as `death in prison'. In the remaining scenarios, we assume that the person is released from jail after serving the entire sentence. In all scenarios, we assume that the person rejoins the cartel upon release. People who are released from prison are then exposed to risks of being killed, reincarcerated, or might retire. We measure the frequency of each scenario. To account for the variability induced by randomly assigning a sentence to a person, we repeat the assignment of sentences to incapacitated individuals. By considering 100 repetitions, we obtain the observed departures that may arise from random sentence assignment. 
}

\subsection{Cartel potential of an age cohort}

%%%% are engaged with 
{
Cartels have infiltrated many economic activities in the continent, including the trade of illicit drugs and weapons \cite{Garcia-PonceHowdoesdrug2021}, of stolen fuel \cite{jones2019huachicoleros}, wildlife, minerals, and even lime and avocados \cite{henkin2020pits}, extortion, kidnapping, and more. To estimate the potential of an age cohort to engage in cartel activities and the amount that is averted through incarceration, the combined number of years that each person could contribute to the organisation from their recruitment to their expected natural death is computed. Under certain assumptions, the number is proportional to the expected severity of crimes committed by cartel members while they are part of the organisation (details in the SI). Measuring cartel activity solely by time in the group overlooks the specialisations of its members \cite{britt1996measurement}, criminal career trajectories \cite{campedelli2021life}, and variation in the intensity of offending \cite{avi1973quantitative}. Yet, the indicator helps obtain estimates of the impact of incarceration and of rehabilitation.
}

%%% the 1990
{
The recruited members of the 1990 cohort could contribute 670,000 years of cartel activities if they all reached their end stage. Then, we quantify how much of that potential was lost by examining the time spent in jail and the dates at which some members were killed and at which they retired. To quantify the cartel activities averted by incarceration, we sum the lengths of the sentences. Of the 670,000 years of potential cartel activities, we obtain that only 46\% are suffered, where a large part is averted due to homicides, then retirement, and finally, due to incarceration.
}

%%% alpha incarc
{
\paragraph{The effect of longer sentences.} To quantify the impact of longer sentences, we follow the same process, but for the incarceration of individuals, we multiply the time assigned for each case by a factor $\alpha \geq 1$. For a specific value of $\alpha$, we determine the years spent in prison, as well as the frequency of death in prison or retirement and homicide upon release. Similarly, if the combined age and length of the sentence exceed the person's expected life, we label the case as death in prison. 
}

%%% effect of rehab
{
\paragraph{The effect of rehabilitation.} To quantify the impact of social rehabilitation, we take a share $\beta \geq 0$ of released individuals and consider that they do not return to the cartel. For those individuals, rather than contributing more time to the effective cartel potential, we remove them from the group (effectively acting as an early retirement). With $\beta = 0$, no rehabilitation occurs, whereas with $\beta = 1$, nobody released from jail returns to the cartel. With complete social rehabilitation, the impact of prison could expand from reducing 9\% of the cartel potential to 29\%.

}

\subsection{Data Availability}

The data used in this study are openly available from INEGI (Instituto Nacional de Estadística y Geografía) \cite{CNSIPEE, inegiPenitenciario, InegiMortalidad, ENPOL}.

\subsection{Code Availability}

The code employed in this study is publicly accessible \href{https://github.com/rafaelprietocuriel/IncarcerationAndRehabilitation}{\textit{here}}. 

%https://github.com/rafaelprietocuriel/IncarcerationAndRehabilitation

\section*{Acknowledgements}

I am grateful to Prof Sergio Aguayo for his careful reading of an earlier version of the manuscript and his insightful comments.

\section*{Author contributions}

The author is solely responsible for all content of this paper.

\section*{Competing interests}

The author declares that they have no competing interests.

\clearpage

\renewcommand{\figurename}{Supplementary Figure}
\renewcommand{\tablename}{Supplementary Table}
\setcounter{table}{0} 
\setcounter{figure}{0}

\section*{Supplementary Information}

\subsection*{A - Age and gender focus}

% why males and why that age
{
Fertility in Mexico has dropped considerably since 1960, from a rate of over six children born per woman to less than 2.0 after 2020. Due to declining fertility rates, one of the largest age cohorts in the country is the 1990 birth cohort, with approximately 1.35 million males. The 1990 cohort in Mexico provides cartels with one of the largest pools of young male recruits \cite{juarez2022significance}. Additionally, the 1990 cohort is relevant to study since they have lived since their teenage years in a country with a war against organised crime cartels. In 2006, Mexico engaged in a complex, violent, and long-lasting fight between the state and organised crime cartels. The Federal Police, a decentralised body of the Ministry of the Interior, was established in 2009 as an operational arm in the fight against organised crime. That cohort has been exposed during their teenage years to cartel recruitment, then, to the formation of the Gendarmeria (during the 2012-2018 government), to the change in policies, to the strategy of `abrazos no balazos' (more hugs and fewer bullets) imposed in the country in 2018, and the militarisation of the country since then. Yet, with changing security policies and strategies, there has been a substantial increase in homicides in the country \cite{InegiMortalidad}, combined with a worrying trend of missing people who are yet to be found \cite{RNPDNO}. Thus, the 1990 cohort has been exposed to the increasing presence of cartels and the waves of violence in the country. 
}

%%% entry and exit. % what if not 35y
{
\paragraph{Modelling the age of recruitment} Determining the lower age for cartel entry, $l$, (i.e., the minimum age for cartel recruitment) and the upper boundary at which recruitment stops because individuals are considered too old, $u$, is not straightforward. Some job listings published on social media to recruit cartel members suggest that cartels often focus their recruitment strategies on young males \cite{CN25000, CJNG25000}. Based on these hints regarding cartel recruitment, we assume $l = 15$ years for cartel entry and $u = 35$ years for the maximum recruitment age. We then sample an age $a$ for each person using the distribution $\phi(a) = s ( a -l) (u -a) I_{[l,u]}(a)$, where the value of $s$ guarantees that the function $\phi(a)$ is a distribution, and $I_{[l,u]}(a)=1$ only for values of $a$ inside the $[l,u]$ interval. 
}

%%%% impact of $m$ and $M$
{
To quantify the impact of assigning two values, $l$ and $u$, to the age distribution of cartel recruitment, we take as maximum age the expression $u + \delta$, for some value of $\delta \in (-5,5)$, so varying the maximum age by assigning values between 30 and 40 years old. With smaller values of $u + \delta$, more individuals are recruited at an early stage, whereas with larger values of $u + \delta$, individuals are more often recruited at later stages. We then assign ages to all recruited individuals and count those who would be considered part of the 1990 age cohort. Results show only a slight variation in the number of members from the 1990 cohort, with differences of less than 8\% compared with the 15,000 to 16,000 members initially obtained (Supplementary Figure \ref{UncertaintyAge}).  
\begin{figure}[!htbp]
\centering
\includegraphics[width=0.5\textwidth]{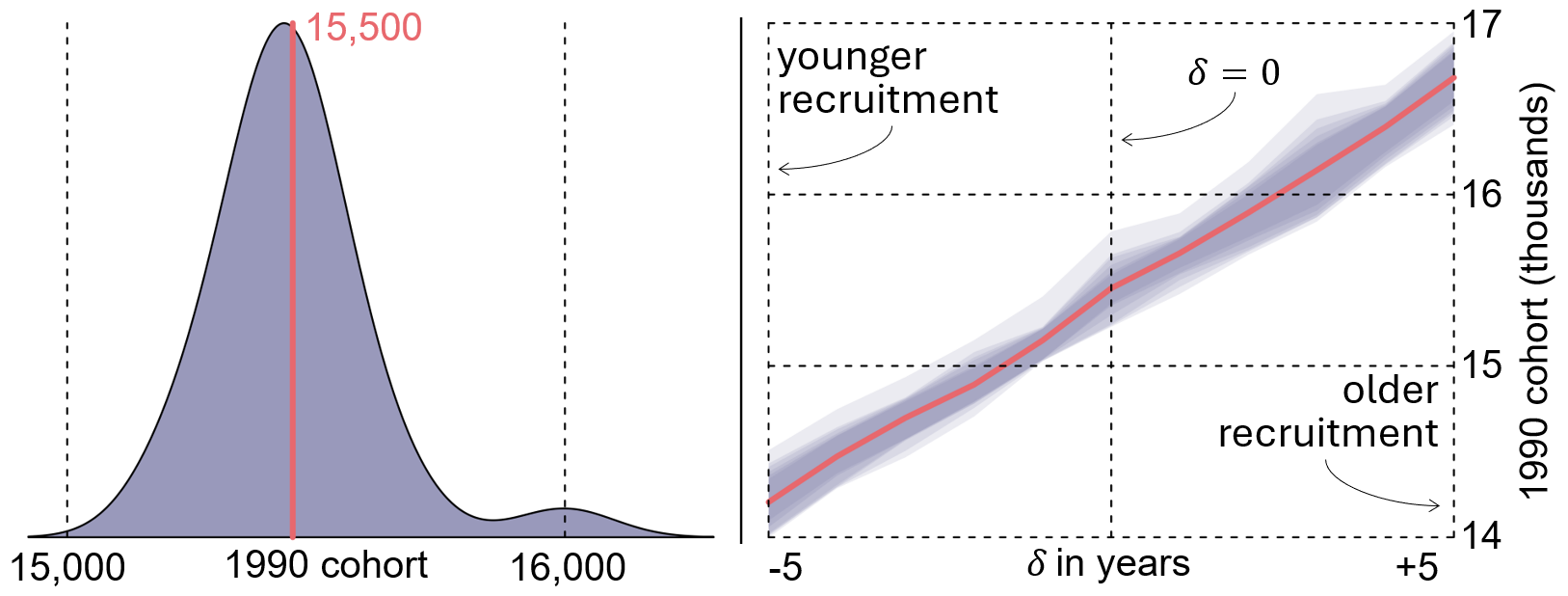}
\caption{Number of recruited members of the 1990 age cohort (vertical axis) when the maximum age for recruitment is defined as $u+\delta$ (horizontal axis). By repeating the same stochastic process, we observe departures in the number of active members across different age-at-recruitment distributions. For values of $\delta =0$, we obtain departures based on the original distribution, mostly between 15,000 and 16,000 members (right panel).}\label{UncertaintyAge}
\end{figure}
}

%%%% reduce then $l$
{
A similar thing happens if, instead of varying the upper age for recruitment, $u$, we focus on the lower age $l$. If we consider the lower bound as $l+\delta$, the 1990 cohort contributes a similar number of members to the cartels, but their contributions occur slightly later. In the extreme case where cartels only recruit people of a specific age (for example, only 18-year-old males), a cohort contributes to all recruited members during the year in which they are that age. The variations in terms of the number of people recruited by the cartel (as observed in the trend in Supplementary Figure \ref{UncertaintyAge}) are not because the dynamics are altered but result only from the increasing number of recruits observed between 2012 and 2022 \cite{PrietoCampedelliHope2023}. Thus, we estimate that between 15,000 and 16,000 members of the 1990 cohort will participate in cartel activities, and similar numbers are observed across all age cohorts between 1986 and 1994 (Supplementary Table \ref{OneMillionTable}).
}

\begin{table}[ht]
\centering
\caption*{Cartel participation in thousands} 
\begin{tabular}{lccc}
  \hline
Cohort & Recruited & Killed & Incapacitated \\ 
  \hline
G-1980 & 11.20 & 4.79 & 4.16 \\ 
  G-1981 & 11.58 & 4.87 & 4.32 \\ 
  G-1982 & 11.93 & 5.15 & 4.36 \\ 
  G-1983 & 12.57 & 5.31 & 4.62 \\ 
  G-1984 & 12.69 & 5.38 & 4.56 \\ 
  \hline
  G-1985 & 13.28 & 5.52 & 4.94 \\ 
  G-1986 & 13.55 & 5.68 & 4.84 \\ 
  G-1987 & 13.99 & 5.82 & 5.02 \\ 
  G-1988 & 14.36 & 5.99 & 5.21 \\ 
  G-1989 & 14.88 & 6.10 & 5.32 \\ 
  \hline
  G-1990 & 15.43 & 6.28 & 5.52 \\ 
  \hline
  G-1991 & 15.77 & 6.47 & 5.66 \\ 
  G-1992 & 16.41 & 6.63 & 5.84 \\ 
  G-1993 & 16.86 & 6.86 & 5.99 \\ 
  G-1994 & 17.18 & 6.90 & 6.11 \\ 
  G-1995 & 18.04 & 7.18 & 6.38 \\ 
  \hline
  G-1996 & 18.30 & 7.33 & 6.31 \\ 
  G-1997 & 19.06 & 7.56 & 6.48 \\ 
  G-1998 & 19.39 & 7.54 & 6.67 \\ 
  G-1999 & 19.94 & 7.68 & 6.54 \\ 
  G-2000 & 20.60 & 7.76 & 6.73 \\
   \hline
\end{tabular}
    \caption{Number of people in thousands recruited, killed, and incapacitated (without considering reincarceration) by cohort.} \label{OneMillionTable}
\end{table}

%%% retirement
{
\paragraph{Modelling the age of retirement.} During week $t$, the model produces two main outputs: the number of people who will be recruited $R(t)$, and the number of active members $A(t)$, those who will be killed $K(t)$, incapacitated $I(t)$, and retired $S(t)$. Each active member has an associated age, represented as the list ${a_1, a_2, \dots, a_{A(t)}}$. To update the list of active members, $K(t)$ individuals are randomly selected and removed to represent those killed, and similarly, $I(t)$ are randomly removed to represent those incapacitated. Retirement, however, depends on age. A simple approach would be to sort the members by age and remove the oldest $S(t)$ individuals, ensuring that the remaining active population stays within a realistic age range. To make this process more realistic and consider other stages for retirement, the probability that an individual $i$ with age $a_i$ retires is defined as $\pi_i = \gamma a_i^\mu$, where $\gamma = 1 / \sum_i a_i^\mu$ ensures that the probabilities sum to one. The parameter $\mu$ controls how strongly age influences the likelihood of retirement. With $\mu = 0$, all individuals have the same probability of retiring (similar to those who are killed or incapacitated), but with larger values of $\mu$, the probability that older members retire increases. For example, if two individuals $i$ and $j$ have ages such that $a_i = 2a_j$ (for example, with 20 and 40 years, respectively), then the probability that $i$ retires is $2^\mu$ times higher than that of $j$ in that week.

To calibrate the value of $\mu$, the following procedure was conducted. For a given value of $\mu$, the model was run by creating a list of active members and updating it each week (accounting for homicides, incapacitations, and retirements) over 80 years. The age distribution of the retired individuals was then analysed. The list contains over 110,000 agents, each with their age at retirement. If the list includes individuals older than 65 years (the country's retirement age), the value of $\mu$ is increased by 1 unit, and the process is repeated. The goal is to identify the smallest possible value of $\mu$ such that the retirement age of all simulated agents remains within a reasonable range. The process stops at $\mu = 15$, which is then used for the analysis. In this stochastic process, most people retire at a certain age, but there is a (rare) chance of retiring earlier.
}

\subsection*{B - The penitentiary system in Mexico}

%%%%
{
The Mexican Federal Penal Code defines organised crime in the ``Ley Federal Contra la Delincuencia Organizada'' (Federal Law Against Organised Crime) when three or more persons organise to commit crimes like terrorism, drug-trafficking, illicit arms trafficking, and others \cite{codigoPenalMexicano}. The penalty for organised crime ranges from 4 to 40 years, depending on the severity of the crime and whether the person held leadership roles \cite{LFCDO}. Data regarding people who entered prison is available for different age groups in Mexico \cite{CNSIPEE}. However, linking organised crime with incarcerations is not methodologically possible because of many reasons, including the type of crime for which cartel members may be sentenced. Data from the largest survey of the country's prison population indicates the length of their sentences \cite{ENPOL}. In total, 40,000 cases are considered. The distribution yields an average sentence length of approximately 20 years for those surveyed (Supplementary Figure \ref{SentenceLength}). However, the observed distribution does not correspond to the frequency with which sentences are assigned. In general, the frequency with which a sentence is assigned is proportional to the observed frequency divided by the length of that sentence (as detailed below with an example). 
\begin{figure}[!htbp]
\centering
\includegraphics[width=0.5\textwidth]{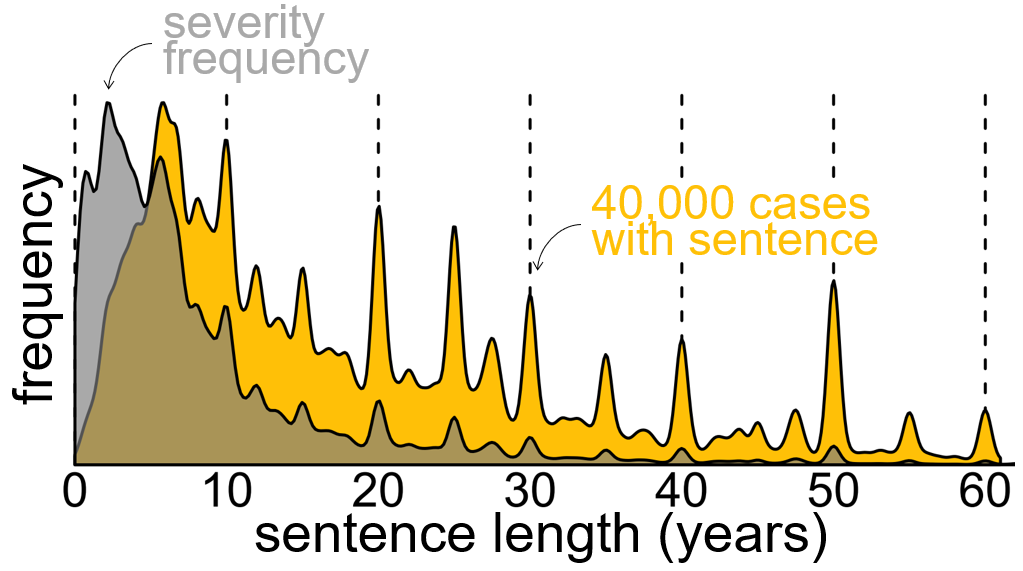}
\caption{Observed distribution of sentences in prison based on the prison population survey \cite{ENPOL}. Utilising that survey, we approximate the frequency with which each sentence is assigned. That distribution is then used to analyse time spent in prison.}\label{SentenceLength}
\end{figure}
}

%%% example of short and long
{
Consider the scenario in which sentences are either 1 year or 20 years in prison. If we were to observe 50 people with a 20-year sentence and 50 people with a one-year sentence, what would be the frequency with which each sentence is assigned? Since the date in which the survey is conducted is independent of the sentences being experienced in prison, we get that the probability of observing the longer sentence (say $\pi_{20}$) and the probability of a shorter sentence (say $\pi_{1}$) are not the same since a person with a longer sentence is more likely part of the surveyed population. The ratio of the probabilities is proportional to the ratio of their lengths, so we get that $\pi_{20} / \pi_{1} = 20$. From this, the observed number of 20-year sentences should be $N f_{20} \pi_{20}$, where $N$ is the number of observed cases, and $f_{20}$ is the frequency with which a judge assigns the 20-year sentences (and similarly, $f_1$ is the frequency of 1-year sentences). It follows that the ratio of the observed frequencies must be equal to $f_{20} \pi_{20} / f_{1} \pi_{1} = 20 f_{20} / f_{1}$. Thus, if we observe equal frequencies of 20- and 1-year sentences (as in the example), then for every 20-year sentence, we should observe 20 1-year sentences. 
}

% repeat the sampling process 1m times
{
We use the frequency of observed sentence lengths to assign them to incarcerated cartel members \cite{ENPOL}. However, we apply the principle that longer sentences are more likely to be observed in a survey to randomly assign sentences to incarcerated individuals. By assigning a sentence to each case and considering the frequency with which they are assigned, we create a realistic scenario of sentence length in the country. On average, sentences assigned to incarcerated people are eight years long. Furthermore, by repeating the same process multiple times (in our case, 100 times), we can consider the effect of assigning random sentences to different individuals. What changes between simulations is the length of the sentence (so a young person could receive a short sentence and rapidly return to cartels in one simulation, but receive a long sentence in the next).
}

%%%% observed recidivism
{
\paragraph{Reincarcerated offenders.} Data from incarcerations in Mexico reports whether the person already had a criminal record \cite{CNSIPEE}. Between 2020 and 2024, over 580,000 males were incarcerated, of whom approximately 22\% had a criminal record (Supplementary Table \ref{RecidTable}). This rate is considerably lower for the 53,000 females who were incarcerated, of which 12\% had a criminal record. For 2020 and 2021, the incarceration and reincarceration rates are lower than those in subsequent years, likely due to changes related to the COVID-19 pandemic \cite{BalmoridelaMiyarDruglordsdonstay2021, GomezOrganisedCrimeGovernance2020}.

\begin{table*}[ht]
\centering
\caption*{Incarcerations with previous criminal record} 
\begin{tabular}{lccc|ccc}
 & \multicolumn{3}{c}{males} & \multicolumn{3}{c}{females} \\
 & incarcerations & with record &   & incarcerations & with record &  \\
year & (thousands) & (thousands) & (\%) &  (thousands) & (thousands) & (\%)\\
  \hline
2024 & 124.4 & 30.5 & 24.5 & 12.2 & 1.8 & 15.1 \\ 
2023 & 132.3 & 27 & 20.4 & 12.5 & 1.4 & 11.3 \\ 
2022 & 125.8 & 28.4 & 22.6 & 11.2 & 1.4 & 12.7 \\ 
2021 & 105.1 & 23.1 & 22 & 9.3 & 1 & 10.3 \\ 
2020 & 94.8 & 16.5 & 17.4 & 7.9 & 0.7 & 8.5 \\ 
\hline
2020-2024 & 582.3 & 125.5 & 21.6 & 53.0 & 6.3 & 11.9 \\ 
\end{tabular}
\caption{Incarcerations in Mexico by gender per year, considering those with a previous criminal record. Data from \cite{CNSIPEE}. } \label{RecidTable}
\end{table*}

}

\subsection*{C - Demography in Mexico under the exposure of cartels}

%%% add details on births, fertility and life expectancy at birth for 69 years
{
The life expectancy at birth for people from the 1990 cohort in Mexico was estimated at approximately 69 years \cite{UNPopulationData}. That number indicates the number of years a newborn would live if prevailing patterns of mortality at the time of its birth were to stay the same throughout its life. However, patterns of mortality are substantially different due to organised crime. Results show that many cartel members are victims of homicide, and they are, on average, 33.6 years old. The life expectancy of the 16,000 males recruited by a cartel decreases by 14.4 years compared to those who were not recruited, from 69 to 54.6 years of life.
}

\subsection*{D - Crimes averted during prison}

%%% model crime intensity
{
How to model criminal careers and how much crime is avoided by incarceration are questions that have been studied for decades \cite{glueck1950unraveling}. One way to model this is to examine the intensity of crimes each person commits. Let $X_i(t)$ denote the number of crimes that offender $i$ commits from some initial period $t=0$ until time $t$. This process can be modelled with a Poisson distribution, where the rate $\lambda_i$ represents the criminal \textit{intensity} of that person \cite{MixedPoissonNagin}. This formulation captures the idea that some offenders are more productive (higher intensity) than others, corresponding to a higher value of $\lambda_i$. The individual is expected to commit $\lambda_i t$ crimes during that period.   
}

%%%% multinomial
{
Once a person has decided to commit a crime, we model the type of crime they will commit. Let $\mathcal{J}$ denote the set of all possible crimes in which a person can be involved (for example, extortion, car theft, kidnapping, or others). Among these options, the individual selects the crime type $j$ with probability $\pi_j$, where $\sum_{j \in \mathcal{J}} \pi_j = 1$. The process of selecting the type of crime is modelled with a Multinomial distribution $\text{Multin}(\pi)$, where $\pi = (\pi_1, \pi_2, \dots, \pi_n)$ represents the probability of each crime type. 
}

%%% mixt
{
By combining the intensity and type of crimes, we obtain that individual $i$ commits crime $j$, expressed as $\mathcal{X}_{ij}(t) \sim \text{Pois}(\pi_j \lambda_i t)$, following a Poisson distribution. There are many options for aggregating different types of crime into a single indicator, such as the financial implications or the penalties imposed upon conviction. Let $w_j$ denote the \textit{severity} associated with committing crime $j$. The purpose of defining the severity of each type of crime is to obtain a comparable index among them. For cartel-related crimes, we can use the sentence length for each crime type. Thus, for different types of crime, severity can be represented by the expected time a person would spend in prison if found guilty. For example, the Penal Code establishes a sentence between 5 and 15 years for threats of terrorist acts and between 15 and 40 years for terrorist acts \cite{codigoPenalMexicano}. To simplify, we define the severity $w_j$ of crime type $j$ as the midpoint of the incarceration range, so the severity of a terrorist threat, for example, would be 7.5 units, while that of a terrorist act would be 27.5 (Supplementary Table \ref{penal_code_severity}).

\begin{table*}[h!]
\centering
\caption*{Selected crimes and corresponding prison sentences from the Mexico City Penal Code (2025).}
\begin{tabular}{lccc}
\hline
\textbf{Crime} & \textbf{Article} & \textbf{Prison sentence (years)}& \textbf{Severity $w$} \\ 
\hline
Kidnapping & 163 & 40 -- 60 & 50 \\
Kidnapping of a minor & 166 & 50 -- 70 & 60\\
Kidnapping causing death & 165 & 50 -- 70 & 60\\
Express kidnapping & 163(b) & 20 -- 40 & 30\\
Extortion & 236 & 10 -- 15 & 12.5\\
Dispossession of property or land & 238 & 6 -- 10 & 8\\
High-value corruption by a public servant & 256 & 10 -- 25 & 17.5 \\
Illicit enrichment by a public servant & 224 & 2 -- 14 & 8 \\
Rape & 265 & 8 -- 20 & 14 \\
Robbery & 220 & 0.5 -- 10 & 5.25\\
Robbery with violence & 225 & 2.5 -- 18 & 10.25\\
\hline
\end{tabular}
\caption{The severity of a crime is one of the options in which different types can be aggregated and compared, using the midpoint of the incarceration range.} \label{penal_code_severity}
\end{table*}
}

%%%% combine
{
By combining the severity of each crime type with its intensity, we obtain that over the period $t$, an offender is expected to generate a total severity of
\begin{equation}
W_i(t) = \sum_{j \in \mathcal{J}} w_j \left( \pi_j \lambda_i t \right) = \lambda_i t \sum_{j \in \mathcal{J}} w_j \pi_j.
\end{equation}
The expected severity can therefore be expressed as $\gamma_i t$, where $\gamma_i = \lambda_i \sum_{j \in \mathcal{J}} w_j \pi_j$. This indicates that the expected severity increases linearly with the time spent in the cartel. Also, we could use other indicators of severity, such as the financial and non-financial consequences of criminal behaviour, among other measures \cite{cohen2020costs}. 
}

%%% caveats
{
Modelling crime intensity with a constant rate simplifies the treatment of member heterogeneity and offending trajectories \cite{piquero2008taking}. This assumption is particularly relevant because organised crime groups often exhibit a division of labour based on specialisation and specific roles \cite{campedelli2021life}, and co-offending increased with years of criminal experience \cite{meneghini2022co}. Despite these simplifications, the framework allows us to quantify the harm generated by the time members spend within the group.
}

%%% sum and cartel potential
{
By combining the severity, we model the cartel's overall potential as a function of the total time spent by all members. The underlying principle is that there is heterogeneity in the harm inflicted by individual members (captured by the different values of $\gamma_i$ across offenders), but the primary determinant of the cartel's potential can be quantified by the cumulative time its members spend within the organisation.
}

\subsection*{E - The size of the prison population}

%%% not all
{
The model presented is not intended to determine the size of a country's prison population, since many incarcerated people are not part of a cartel. Thus, the model only considers a portion of the country's prison population, consisting only of those who were affiliated with a cartel. 
}

%%% 
{
By examining the period during which cartel members are incarcerated and their jail time, it is possible to obtain two estimates regarding the cartel members currently in prison. First, over 6,000 cartel members will be detained in 2025, among the more than 120,000 incarcerations that will occur that year (Supplementary Table \ref{RecidTable}). 
}

%%% how many there are
{
Second, considering the length of the sentence that each person receives, approximately 52,000 cartel members are in prison today. That is approximately 22\% of the incarcerated population of the country \cite{CNSIPEE}.
}

\subsection*{F - Relapse of criminal behaviour}

%%%5 intro{
{
Recidivism is a broad term that refers to the relapse of criminal behaviour \cite{butorac2017challenges}. It usually includes many outcomes, such as re-arrest, reconviction, and reimprisonment, and is typically measured by crimes that led to those outcomes \cite{butorac2017challenges}. 
}

%%% recidivism is used
{
The relevance of examining recidivism rates is that they are commonly used as proxies for quantifying whether an offender committed a subsequent offence \cite{jennings2019recidivism}. Thus, the rate is often used to quantify the effectiveness of sanctions for preventing future offences. 
}

%%% indicators
{
A common indicator of recidivism is the 2-year reconviction rate \cite{fazel2015systematic}. Across many countries, the 2-year reconviction rate ranged from 14\% to 43\% in men and from 9\% to 35\% in women \cite{yukhnenko2019recidivism}. However, comparing recidivism across countries might be misleading, as reported rates are sensitive to several measurement variables, including definitions, follow-up length, and type \cite{yukhnenko2023criminal}. Additionally, this number may be misleading if past offenders are not adequately traced and if the impunity rate is high. For cartel members, we assume that all individuals return to the group upon release. And even under such conditions, due to the large number of members that are never incapacitated, we obtain that only 25\% of individuals who have been imprisoned once and released will be reincarcerated. Thus, observing a reincarceration rate of 25\% could be interpreted as evidence of some effectiveness of incarceration in preventing future cartel offences. However, this could also be the result of the negligible effectiveness of incarceration combined with high impunity, as observed in the country.
}

\subsection*{G - Social reintegration programs in Mexico}

%%% some social programs
{
Some social reintegration programs work with penitentiary populations in Mexico. For example, \href{https://reinserta.org/}{\textit{Reinserta}} is a nonprofit organisation working with people affected by the penitentiary system. Their programs support children who live in prison with their incarcerated mothers, youth in conflict with the law, and women in prison. Through some targeted interventions, they aim to create a long-term social impact. \href{https://lacana.mx/}{\textit{La Cana}}, by contrast, is a social enterprise explicitly focused on supporting women in Mexican prisons through dignified employment and skill development. They promote fair working conditions and provide follow-up support upon release, helping women transition into stable jobs and reduce the risk of reoffending.
}

%%%% limited
{
However, although some programs work with the prison population, many jails are overpopulated and have limited effects on reintegration \cite{lessing2017counterproductive}. The challenge is to scale efforts such as Reinserta or La Cana to the nearly 150,000 people who are incarcerated each year and will eventually be released \cite{CNSIPEE}.
}

\end{document}